\documentclass[aps,pra,onecolumn,showpacs,amsmath,amssymb,floatfix,superscriptaddress,notitlepage]{revtex4-1}

\usepackage{color}
\usepackage{bbm}
\usepackage{graphicx}
\usepackage{dcolumn}
\usepackage[utf8]{inputenc}
\usepackage{epstopdf}
\usepackage{amsthm}
\usepackage{amsmath}
\usepackage{empheq}
\usepackage{bbm}
\usepackage{braket}
\usepackage{amssymb}
\usepackage[usenames,dvipsnames]{xcolor}
\usepackage{cases}
\usepackage{latexsym}
\usepackage[colorlinks=true,citecolor=Cerulean,linkcolor=RubineRed,urlcolor=Cerulean]{hyperref}

\newcommand{\LP}[1]{{\color{black}  #1}}

\graphicspath{ {images/}{images/Figs-07-23-18/}{images/figure1/}{images/figure1/finales/}{images/figure2/}{images/figure3/}{images/figure3/finales/}{images/figure4/}{images/figure5/} }

\graphicspath{ {images/}{images/Figs-07-23-18/}{images/figure1/}{images/figure1/finales/}{images/figure2/}{images/figure3/}{images/figure3/finales/}{images/figure4/}{images/figure5/}{images/figure1SM/}{images/figure2SM/}}

\newcommand{\id}{\mathbbm{1}} 
\newcommand{\tr}[1]{\operatorname{\textnormal{Tr}}\left( {#1} \right)}  
\newcommand{\rhog}[1]{ {\rho_{#1}^\mathcal{O}}} 
\newcommand{\rhoa}[1]{ {\rho_{#1}^\mathcal{A}}} 
\newcommand{\rhob}[1]{ {\rho_{#1}^\mathcal{B}}} 
\newcommand{\dist}{\mathcal{D}} 
\newcommand{\deph}{{\Lambda}} 

\newcommand{\purityA}{ \mathcal{P} \left( \rhoa{T} \right) }
\newcommand{\purityB}{ \mathcal{P} \left( \rhob{T} \right) }

\usepackage{lipsum}
\usepackage{graphicx}

\begin{document}
\title{ Limits to Perception by Quantum Monitoring with Finite Efficiency }
\author{Luis Pedro Garc\'ia-Pintos}
\affiliation{Department of Physics, University of Massachusetts, Boston, Massachusetts
02125, USA}
\affiliation{Joint Center for Quantum Information and Computer Science and Joint Quantum Institute, NIST/University of Maryland, College Park, Maryland 20742, USA}

\author{Adolfo del Campo}
\affiliation{Department  of  Physics  and  Materials  Science,  University  of  Luxembourg,  L-1511  Luxembourg, Luxembourg}
\affiliation{Donostia International Physics Center,  E-20018 San Sebasti\'an, Spain}
\affiliation{IKERBASQUE, Basque Foundation for Science, E-48013 Bilbao, Spain}
\affiliation{Department of Physics, University of Massachusetts, Boston, Massachusetts
02125, USA}
\date{\today}

\begin{abstract}

We formulate limits to perception under continuous quantum measurements by comparing the  quantum states assigned by agents that have partial access to measurement outcomes.
To this end, we provide bounds on the trace distance and the relative entropy between the assigned state and the actual state of the system. These bounds are expressed solely in terms of the purity and von Neumann entropy of the state assigned by the agent, and are shown  to characterize how an agent's perception of the system is altered by access to additional information. We apply our results to Gaussian states and to the dynamics of a system embedded in an environment illustrated on a quantum Ising chain.

\end{abstract}

\maketitle

Quantum theory rests on the fact that the \emph{quantum state} of a system encodes all predictions of possible measurements as well as the system's posterior evolution. However, in general different agents may assign different states to the same system, depending on their knowledge of it. 
Complete information of the physical state of a system is equated to pure states, mathematically modeled by unit vectors in Hilbert space. In contrast, mixed states correspond to a lack of complete descriptions of the system, either due to uncertainties in the preparation, or due to the system being correlated with secondary systems. 
 In this paper, we address  how  the perception of a system differs among observers with different levels of knowledge.
Specifically, we quantify how different the effective descriptions that two agents provide of the same system can be, when acquiring information  through continuous measurements. 


Consider a monitored quantum system, that is, a system being continuously measured in time.
An omniscient agent $\mathcal{O}$ is assumed to know all interactions and measurements that occur to the system. In particular, she has access to all outcomes of measurements that are performed. 
 As such, $\mathcal{O}$ has a complete description of the \emph{pure state} $\rhog{t} = \left(\rhog{t}\right)^2$ of the system.  
 
While not necessary for subsequent results, we model such monitoring process by continuous quantum measurements~\cite{JacobsIntro2006,
Bookwiseman2009,
Bookjacobs2014} as a natural test-bed with experimental relevance~\cite{MurchNature2013,
ScienceDevoret2013,SiddiqiNature2014}.
 For ideal continuous quantum measurements, 
 the state $\rhog{t}$ satisfies a stochastic equation dictating its change,
\begin{align}
\label{eq:master}
d \rhog{t} = -i\left[ H,\rhog{t} \right] dt +  \deph \left[ \rhog{t} \right]dt + \sum_\alpha I_{A_\alpha}\left[ \rhog{t}\right] dW_t^\alpha.
\end{align}
The dephasing superoperator $\deph \left[ \rhog{t} \right]$ is of Lindblad form, 
\begin{align}
\label{eq:lindblad}
\deph\left[ \rhog{t} \right] =   - \sum_\alpha \frac{1}{8 \tau_m^\alpha } \left[A_\alpha, \left[A_\alpha,\rhog{t} \right] \right]
\end{align}
for the set of measured physical observables $\{ A_\alpha \}$, 
and the ``innovation terms" are given by
\begin{align}
\label{eq:innovation}
I_{A_\alpha}\left[ \rhog{t} \right] = \frac{1}{\sqrt{4\tau_m^\alpha}} \left( \{ A_\alpha, \rhog{t} \} - 2\tr{A_\alpha \rhog{t}} \rhog{t} \right).
\end{align}
The latter account for the information about the system acquired during the monitoring process, and model the quantum back-action on the state during a measurement. 
The \emph{characteristic measurement times} $\tau_m^\alpha$ depend on the strength of the measurement, and characterize the time over which information of the observable $A_\alpha$ is acquired. The terms $dW_t^\alpha$ are independent random Gaussian variables of zero mean and variance $dt$.

An agent $\mathcal{A}$ without access to the measurement outcomes possesses a different --incomplete-- description of the state of the system. The need to average over the unknown results implies that the state $\rhoa{t}$ 
assigned by $\mathcal{A}$ satisfies the master equation
\begin{align}
\label{eq:masterAlice}
d \rhoa{t} =-i\left[ H,\rhoa{t} \right] dt +  \deph \left[ \rhoa{t} \right]dt,
\end{align}
obtained from~\eqref{eq:master} by using that $\langle dW_t^\alpha \rangle = 0 $, where $\langle \cdot \rangle $ denote averages over realizations of the measurement process~\cite{JacobsIntro2006}. Assuming that agent $\mathcal{A}$ knows the initial state of the system before the measurement process, $\rhog{0} = \rhoa{0}$, the state that she assigns at later times is $\rhoa{t} \equiv \langle \rhog{t} \rangle$.

As a result of the incomplete description of the state of the system, 
agent $\mathcal{A}$ suffers from a growing uncertainty in the predictions of measurement outcomes.
We quantify this by means of two figures of merit: the trace distance and the relative entropy.

The trace distance \LP{between states $\sigma_1$ and $\sigma_2$} is defined as 
\begin{align}
\label{eq:tracedistance}
\dist(\sigma_1,\sigma_2) = \frac{\| \sigma_1 - \sigma_2 \|_1}{2},
\end{align}
where the trace norm for an operator with a spectral decomposition $A = \sum_j \lambda_j \ket{j} \bra{j}$ is $\| A \|_1 = \sum_j |\lambda_j|$.
Its operational meaning derives from the fact that the trace distance characterizes the maximum difference in probability of outcomes for any measurement on the states $\sigma_1$ and $\sigma_2$:
\begin{align}
\dist(\sigma_1,\sigma_2) = \max_{0 \le P \le \id} | \tr{P\sigma_1} - \tr{P\sigma_2} |,
\end{align}
where $P$ is  a positive-operator valued measure.
It also quantifies the probability $p $ of successfully guessing, with a single measurement instance, the correct state in a scenario where one assumes equal prior probabilities for having state $\sigma_1$ or $\sigma_2$. 
Then, the best conceivable protocol gives $p  = \frac{1}{2}\left( 1 + \dist(\sigma_1,\sigma_2) \right)$.
Thus, if two states are close in trace distance they are hard to distinguish under any conceivable measurement~\cite{Booknielsenchuang,wilde_2013,watrous_2018}.

The relative entropy also serves as figure of merit to quantify distance between probability distributions, in particular characterizing the extent to which one distribution can encode information contained in the other one~\cite{cover2012elements}. 
In the quantum case, the relative entropy is defined as
\begin{align}
S\left( \sigma_1 || \sigma_2 \right) \equiv \tr{\sigma_1 \log \sigma_1} - \tr{\sigma_1 \log \sigma_2}. 
\end{align}
In a hypothesis testing scenario between states $\sigma_1$ and $\sigma_2$, the probability $p_N$ of wrongly believing that $\sigma_2$ is the correct state scales as $p_N \sim e^{-N S\left( \sigma_1 || \sigma_2 \right)}$ in the limit of large $N$, where $N$ is the number of copies of the state that are available to measure on~\cite{Petz1991,Ogawa2005}.
That is, $\sigma_2$ is easily confused with $\sigma_1$ if $S\left( \sigma_1 || \sigma_2 \right)$ is small~\cite{schumacher2002relative,VedralRevModPhys2002}.

\

\section{Quantum limits to perception}
\begin{figure}
  \centering
           \includegraphics[angle=90,origin=c,trim=00 00 00 00,width=0.5\textwidth]{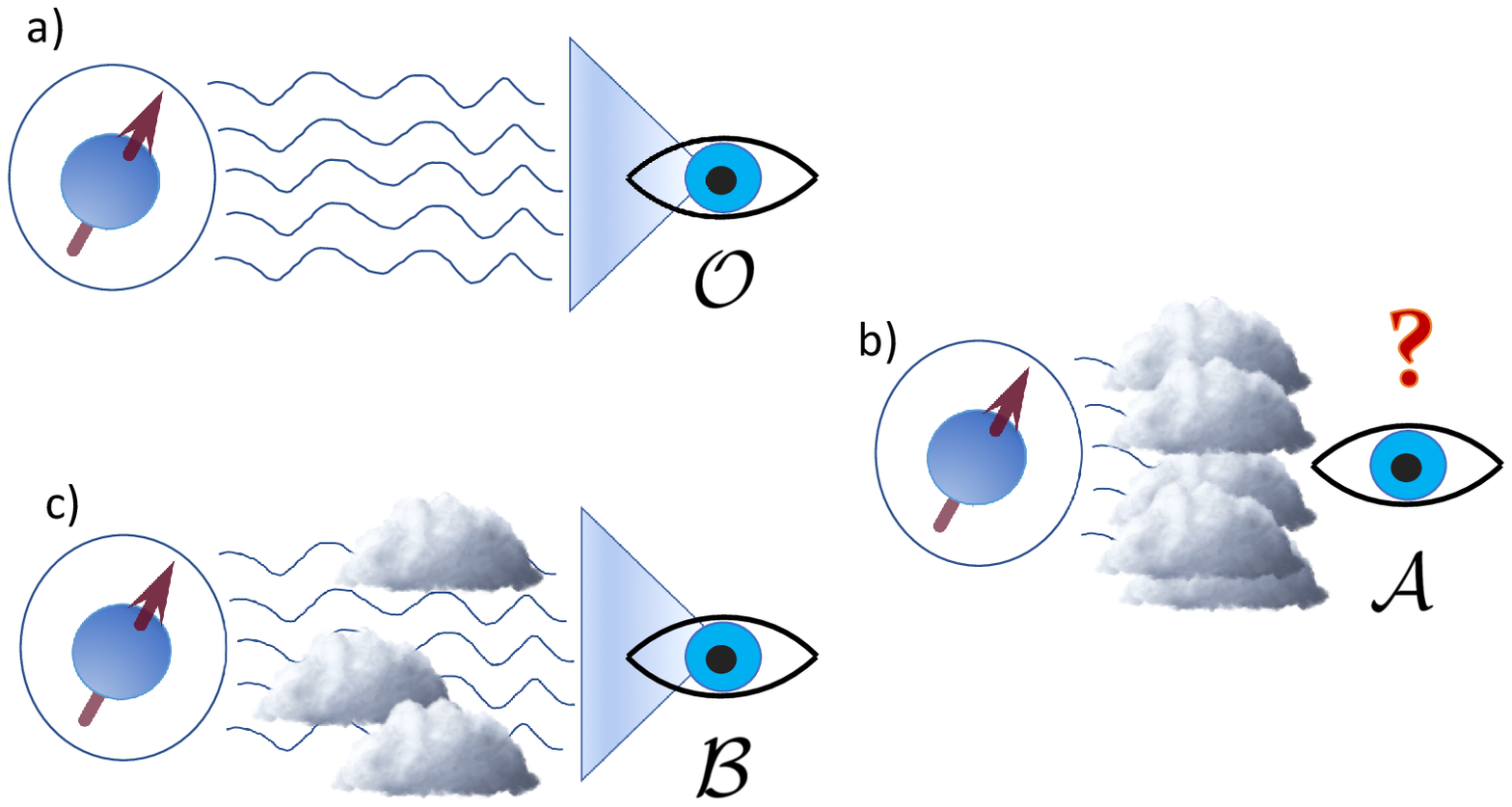}  
\caption{
\label{fig:ObserversPerceptions} 
{\bf
Illustration of the varying degrees of perception by different agents.} 
The amount of information that an agent possesses of a system can drastically alter its perception, as
the expectations of outcomes for measurements performed on the system can differ.
a) The state $\rhog{t}$ assigned by omniscient agent $\mathcal{O}$, who has full access to the measurement outcomes, corresponds to a complete pure-state description of the system. $O$ thus has the most accurate predictive power. 
b) An agent $\mathcal{A}$ completely ignorant of measurement outcomes possesses the most incomplete description of the system.
c) A continuous transition between the two descriptions, corresponding to the worst and most complete perceptions of the system respectively, is obtained by considering an agent $\mathcal{B}$ with partial access to the measurement outcomes of the monitoring process.
}
\end{figure}

Lack of knowledge of the outcomes from measurements performed on the system \LP{induces $\mathcal{A}$ to assign an incomplete, mixed, state to the system. This hinders the agent's perception of the system (see illustration in Fig.~\ref{fig:ObserversPerceptions}).} We quantify this by the trace distance and the relative entropy. 

We are interested in comparing $\mathcal{A}$'s incomplete description to the pure state $\rhog{T}$ assigned by $\mathcal{O}$, i.e., to the complete description.
Under ideal monitoring of a quantum system, the pure state $\rhog{T}$ remains pure. 
Therefore, 
 the following holds~\cite{Booknielsenchuang}
\begin{align}
1  - \tr{\rhog{T} \rhoa{T}} \le \dist\left(\rhog{T},\rhoa{T}\right) \le \sqrt{1  - \tr{\rhog{T} \rhoa{T}} }.
\end{align}
One can then  
 directly relate the average trace distance to the purity $\mathcal{P} \left( \rhoa{T} \right)  \equiv  \tr{\rhoa{T}^2}$ of state $\rhoa{T}$ as
\begin{align}
\label{eq:bound-alice}
1  -  \purityA   \le \left\langle \dist \left(\rhog{T},\rhoa{T}\right) \right\rangle \le \sqrt{1  - \purityA },
\end{align}
by using Jensen's inequality and the fact that the square root is concave.
The level of mixedness of the state $\rhoa{T}$ that $\mathcal{A}$ assigns to the system provides lower and upper bounds to the average probability of error that she has in guessing the actual state of the system $\rhog{T}$. This provides an operational meaning to the purity of a quantum state, as a quantifier of the average trace distance between a state $\rhog{t}$ and post-measurement (average) state $\rhoa{t}$.

To appreciate the dynamics in which  the average trace distance evolves, we note that at short times
\begin{align}
\frac{T}{\tau_D}\le \left\langle \dist \left(\rhog{T},\rhoa{T}\right) \right\rangle \le \sqrt{\frac{T}{\tau_D}},
\end{align}
where the decoherence rate is given by~\cite{Chenu17,Beau17}
\begin{align}
\frac{1}{\tau_D}= \sum_\alpha \tfrac{1}{4\tau_m^\alpha } {\rm Var}_{\rhoa{0}}(A_\alpha),
\end{align}
in terms of the variance ${\rm Var}_{\rhoa{0}}(A_\alpha)$ of the measured observables over  the initial pure state $\rhoa{0}$. 
Analogous bounds can be derived at arbitrary times of evolution for the difference of perceptions among various agents (see Appendix). 

For the case of the quantum relative entropy between states of complete and incomplete knowledge, the following identity holds
\begin{align}
\label{eq:bound-relentropy}
\left\langle  S\left( \rhog{t}  || \rhoa{t}  \right) \right\rangle &= S\left(\rhoa{t}\right),
\end{align}
proven by using that $\rhog{t}$ is pure and that the von Neumann entropy of a state $\sigma$ is $S\left( \sigma \right) \coloneqq - \tr{\sigma \log \sigma }$. Thus, the entropy of the state assigned by the agent $\mathcal{A}$  fully determines the average relative entropy with respect  to the complete description $\rhog{t}$~\footnote{Alternative interpretations to this quantity have been given in~\cite{barchielli2002entropy,
barchielli2009quantum}}.

Similar calculations allow to bound the variances of $\dist \left(\rhog{T},\rhoa{T}\right)$ and of $ S\left( \rhog{t}  || \rhoa{t}  \right) $ as well. The variance of the trace distance, $\Delta \dist_T^2 \equiv \left\langle \dist^2\left(\rhog{T},\rhoa{T}\right) \right\rangle - \left\langle \dist\left(\rhog{T},\rhoa{T}\right) \right\rangle^2 $, satisfies
\begin{align}
\label{eq:boundvariance}
\Delta \dist_T^2 \le \purityA -\purityA^2, 
\end{align}
while for the variance of the relative entropy it holds that
\begin{align}
\label{eq:bound-Var-relentropy}
\Delta S^2\left( \rhog{t}  || \rhoa{t}  \right) &\leq   \tr{\rhoa{t} \log^2 \rhoa{t}}  - S^2\left(\rhoa{t}\right).
\end{align}
The right hand side of this inequality admits a classical interpretation in terms of the variance of the surprisal $(-\log p_j)$ over the eigenvalues $p_j$ of $\rhoa{t}$~\cite{VedralRevModPhys2002}.
We thus find that, at the level of a single realization, the dispersion of the relative entropy between the states assigned by the agents $\mathcal{O}$ and $\mathcal{A}$ is upper bounded by the variance of the surprise in the description of $\mathcal{A}$. 
The later naturally vanishes  when $\rhoa{t}$ is pure, and increases as  the state becomes more mixed.

\

\section{Transition to complete descriptions}
So far we considered the extreme case of comparing the states assigned by $\mathcal{A}$,  who is in complete ignorance of the measurement outcomes, and by an omniscient agent $\mathcal{O}$. 
One can in fact consider a continuous transition between these limiting cases, i.e., as the accuracy in the perception of the monitored system by an agent is enhanced,
as illustrated in Fig.~\ref{fig:ObserversPerceptions}. Consider a third agent $\mathcal{B}$, with access to a fraction of the measurement output. This can be modeled by introducing a filter function $\eta(\alpha)\in[0,1]$ characterizing the efficiency of the measurement channels in Eq.~\eqref{eq:master}~\cite{JacobsIntro2006}. Then, 
the dynamics of state $\rhob{t}$ is dictated by
\begin{align}
\label{eq:master-filtered}
d \rhob{t} = -i\left[ H,\rhob{t} \right] dt +  \deph \left[ \rhob{t} \right]dt + \sum_\alpha \sqrt{ \eta(\alpha) } I_{A_\alpha}\left[ \rhob{t}\right] dV_t^\alpha,
\end{align}
with $dV_t^\alpha$ Wiener noises for observer $\mathcal{B}.$
It holds that $\rhob{t} \equiv \langle \rhog{t} \rangle_\mathcal{B}$, where the average is now over the outcomes obtained by $\mathcal{O}$ that are unknown to $\mathcal{B}$~\cite{JacobsIntro2006}. 
 
Note that the case with null measurement efficiencies $\eta(j) = 0$ gives the exact same dynamics as that of a system in which the monitored observables $\{ A_\alpha \}$ are coupled to environmental degrees of freedom, producing dephasing~\cite{ZurekDecoherence,schlosshauer2007decoherence}. Equations.~\eqref{eq:master-filtered} and~\eqref{eq:master} then correspond to unravellings in which partial or full access to environmental degrees of freedom allow learning the state of the \LP{system} by conditioning on the state observed in the environment. Therefore, knowing how $\dist\left( \rhob{t}, \rhog{t}  \right)$ and $ S\left( \rhog{t}  || \rhob{t}  \right)$ decrease as $\eta$ increases directly informs of how much the description of an open system can be improved by observing a fraction of the environment.
 This is reminiscent of the Quantum Darwinism approach, whereby fractions of the environment encode objective approximate descriptions of the system. While in the Darwinistic framework the focus is on environmental correlations, we focus on the state of the system itself.

The results of the previous section hold for partial-ignorance state $\rhob{t}$ as well,
\begin{subequations}
\label{eq:bound-bob}
\begin{align}
\label{eq:boundDist-bob}
1  -  \purityB   \le &\left\langle \dist \left(\rhog{T},\rhob{T}\right) \right\rangle_\mathcal{B} \le \sqrt{1  - \purityB} \\
\label{eq:relEntropy-bob}
&\left\langle  S\left( \rhog{t}  || \rhob{t}  \right) \right\rangle_\mathcal{B}  = S\left(\rhob{t}\right),
\end{align}
\end{subequations}
Similar extensions are obtained for the variances. 
This allows exploring 
the transition from 
the incomplete description of $\mathcal{A}$, to a complete description of the state of the system 
as $\eta \rightarrow 1$. Note that these results hold for each realization of a trajectory of $\mathcal{B}$'s state $\rhob{t}$, and that if one averages over the measurement outcomes unknown to both agents $\mathcal{A}$ and $\mathcal{B}$, Eq.~\eqref{eq:relEntropy-bob} gives $\langle S\left( \rhog{t}  || \rhob{t}  \right) \rangle = \langle S(\rho_t^\mathcal{B}) \rangle$.

\LP{These results allow to compare the descriptions of different agents that jointly monitor a system~\cite{JacobsIntro2006,RuskovPRL2010,
LPGPPRA2016,MartiPRA2017,LPGPPRA2017}. We show in the Appendix that
\begin{align}
\left| \tr{\rhoa{T}^2} - \tr{\rhob{T}^2} \right| \leq \Big\langle \dist \left(\rhoa{T},\rhob{T}\right) \Big\rangle_{\mathcal{A}\mathcal{B}} 
\leq \sqrt{1  - \tr{\rhoa{T}^2} } + \sqrt{1  - \tr{\rhob{T}^2} }.
\end{align}
The joint monitoring of a system by independent observers has been realized experimentally in~\cite{hacohen2016quantum,ficheux2018dynamics}.
}

\section{Illustrations}

\subsection{Evolution of the limits to perception}
Consider a 1D transverse field Ising model, with Hamiltonian
\begin{align}
\label{eq:spinchainH}
H = -h\sum_j^N \sigma_j^x - J \sum_j^{N-1} \sigma_j^z\sigma_{j+1}^z,
 \end{align}
where $\sigma_j^x$ and $\sigma_j^z$ denote Pauli matrices on the $x$ and $z$ directions, and $\{h,J\}$ denote coupling strengths. 

We study the case of observer $\mathcal{O}$ monitoring the individual spin $z$ components. Equation~\eqref{eq:master} thus governs the evolution of the state $\rhog{t}$, with $\{A_\alpha\} = \{\sigma_j^z\}$. Meanwhile, the state assigned by observers with partial access to measurement outcomes follows Eq.~\eqref{eq:master-filtered}. The case
 $\eta(j) = 0$ gives equivalent dynamics to that of an Ising chain in which individual spins couple to environmental degrees of freedom via $\sigma_j^z$, producing dephasing.
 
\begin{figure}
  \centering
        \includegraphics[trim=00 00 00 00,width=0.36\textwidth]{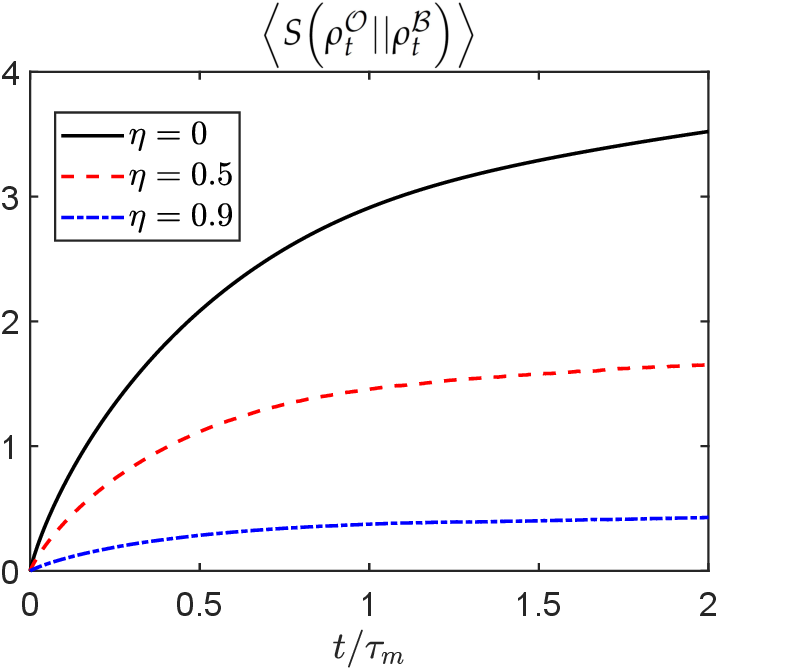}
\caption{
\label{fig:evolutionBounds} 
{\bf Evolution of the average relative entropy.}
Simulated evolution of the average $\left\langle   S\left( \rhog{t}  || \rhob{t}  \right)  \right\rangle = \left\langle   S\left( \rhob{t}  \right)  \right\rangle $ of the relative entropy between complete and incomplete descriptions for a spin chain initially in a paramagnetic state on which individual spin components $\sigma_j^z$ are monitored. Here $\langle \cdot \rangle$ denotes an average over all measurement outcomes, and $\rhob{t} = \langle \rho_t^\mathcal{O} \rangle_\mathcal{B}$ is the state assigned by agent $\mathcal{B}$ after discarding the outcomes unknown to him. 
The simulation corresponds to $N = 6$ spins, with couplings $J\tau_m = h\tau_m = 1/2$.
For $\eta = 0 $ (black continuous curve),  agent $\mathcal{A}$, without any access to the measurement outcomes, has the most incomplete description of the system. 
For $\eta = 0.5$ (red dashed curve), $\mathcal{B}$ gets closer to the complete description of the state of the system, after gaining access to partial measurement results. 
Finally, when $\eta = 0.9$ (blue dotted curve), access to enough information provides $\mathcal{B}$ with an almost complete description of the state. 
Importantly, in all cases the agent can estimate how far the description possessed is from the complete one solely in terms of the entropy $S(\rhob{t})$.
}
\end{figure}

 Figure~\ref{fig:evolutionBounds} illustrates 
the evolution of the averaged relative entropy $\left\langle  S\left( \rhog{t}  || \rhob{t}  \right) \right\rangle$ between the complete description and $\mathcal{B}$'s partial one,
 for different values of the monitoring efficiency $\eta$. The average $\langle \cdot \rangle$ is over all measurement outcomes.  
 Analogous results for the average trace distance can be found in the Appendix.
The dynamics are simulated by implementation of the monitoring process as a sequence of weak measurements, which can be modeled by Kraus operators acting on the state of the system. Specifically, the evolution of $\rhog{t}$ and corresponding state $\rhob{t}$ with partial measurements is numerically obtained from assuming two independent measurement processes, as in~\cite{JacobsIntro2006}.

\color{black}
\

 \subsection{Transition to complete descriptions}
\begin{figure*}
  \centering
        \includegraphics[trim=00 00 00 00,width=0.45\textwidth]{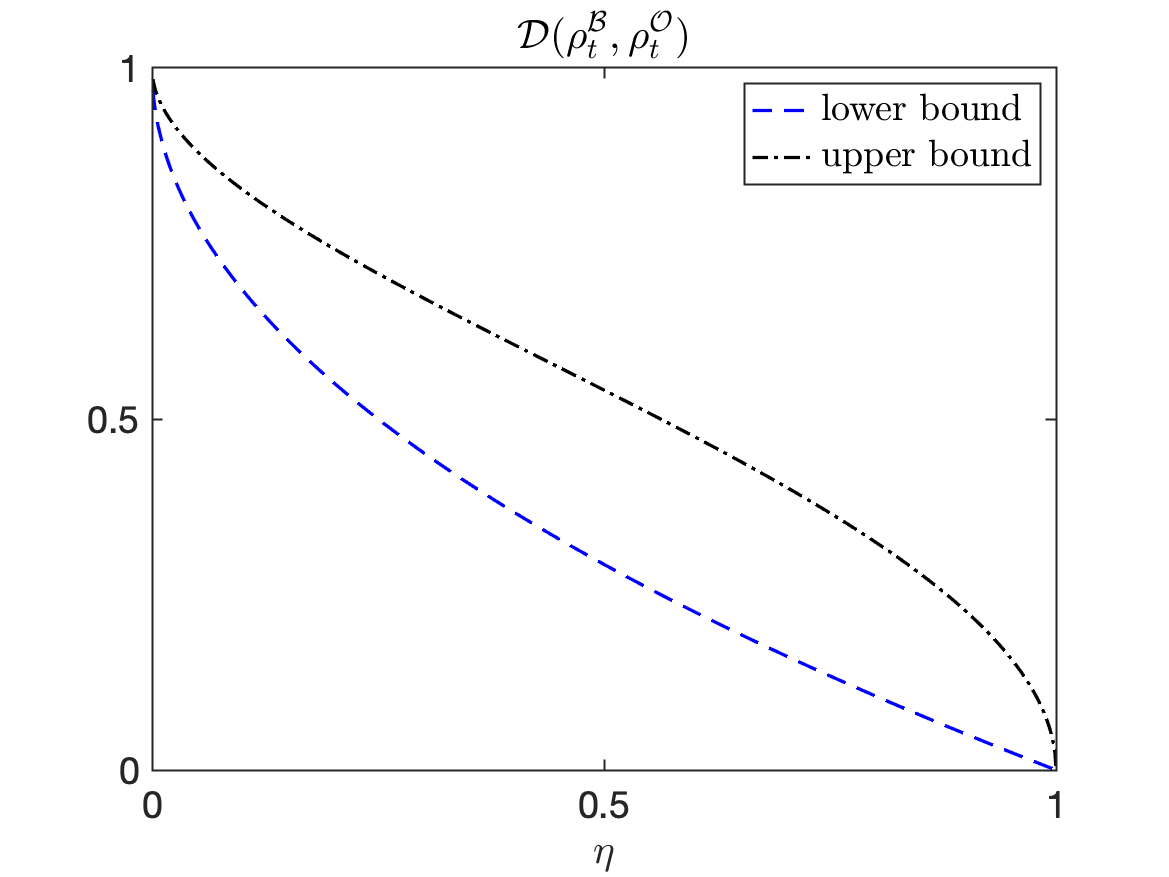}  \hspace{41pt}
         \includegraphics[trim=00 00 00 00,width=0.38\textwidth]{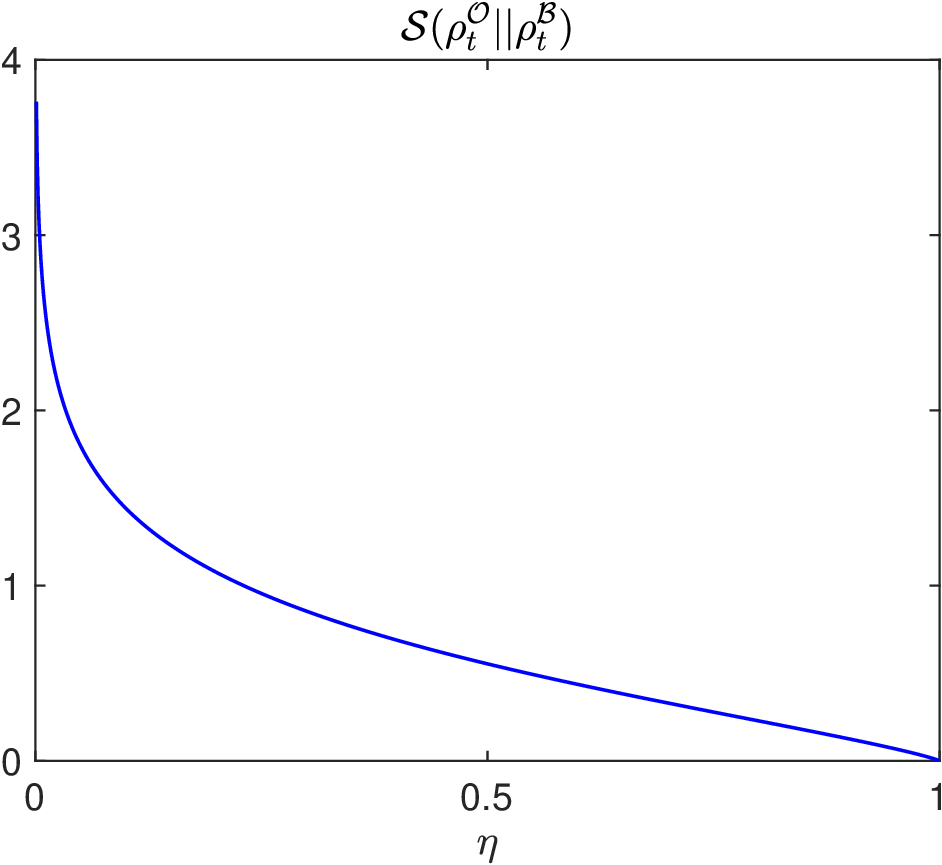}  
\caption{
\label{fig:transition2awareness} 
{\bf 
 Transition between levels of perception.
}
Bounds on average trace distance (left) and average relative entropy (right) as function of measurement efficiency for a harmonic oscillator undergoing monitoring of its position. 
For such system the purity of the state $\rhob{t}$ depends solely on the measurement efficiency with which observer $\mathcal{B}$ monitors the system. This illustrate the transition from complete ignorance of the outcomes of measurements performed ($\eta = 0$), to the most complete description as $\eta \rightarrow 1 $ --- the situation with the most accurate perception. 
Efficient use of information happens when a small fraction of the measurement output is incorporated at $\eta \ll 1$, as then both $  \dist \left( \rhob{t},\rhog{t} \right)  $ and the relative entropy $  S\left( \rhog{t}  || \rhob{t}  \right) $ decay rapidly. 
}
\end{figure*}
Consider the case of a one dimensional harmonic oscillator 
with position and momentum operators $X$ and $P$. 
We assume agent $\mathcal{B}$ is monitoring the position of the oscillator with an efficiency $\eta$. 
The dynamics  
 is dictated by Eq.~\eqref{eq:master-filtered} for the case of a single monitored observable $X$, and can be determined by a set of differential equations on the moments of the Gaussian state $\rhob{t}$~\cite{DohertyPRA99,JacobsIntro2006}.

We prove in the Appendix that the purity of the density matrix for long times has a simple expression in terms of the measurement efficiency, satisfying 
$\purityB   \longrightarrow  \sqrt{\eta}$ for long times. 
Equation~\eqref{eq:bound-bob}
and properties of Gaussian states~\cite{PhysRevA.68.012314,ParisGaussian2005,
HayashiGaussian2007,PirandolaRevModPhys2012,adesso2014continuous}
then imply
\begin{align}
1  -  \sqrt{\eta}   \le \left\langle \dist \left(\rhog{T},\rhob{T}\right) \right\rangle_\mathcal{B} \le \sqrt{1  - \sqrt{\eta} },
\end{align}
and
\begin{align}
\left\langle  S\left( \rhog{t}  || \rhob{t}  \right) \right\rangle_\mathcal{B} &=  \left( \frac{1}{ 2\sqrt{\eta} } + \frac{1}{2} \right) \log \left( \frac{1}{2\sqrt{\eta} } + \frac{1}{2} \right)  -  \left( \frac{1}{ 2\sqrt{\eta} } - \frac{1}{2} \right) \log \left( \frac{1}{2\sqrt{\eta} } - \frac{1}{2} \right) .
\end{align}
 See~\cite{WisemanPRLGaussianSmoothing2019} for further results on the gains in purity that can be obtained from conditioning on measurement outcomes in Gaussian systems.
Figure~\ref{fig:transition2awareness} depicts the trace distance $\left\langle \dist \left( \rhob{t},\rhog{t} \right) \right\rangle_\mathcal{B}$ and the relative entropy $\left\langle  S\left( \rhog{t}  || \rhob{t}  \right) \right\rangle_\mathcal{B}$ as a function of the measurements efficiency of $\mathcal{B}$'s measurement process, illustrating the transition from 
least accurate perception to most accurate perception and
 optimal predictive power as $\eta \rightarrow 1$.
Note that, since both the bounds on the trace distance and relative entropy are independent of the parameters of the model in this example, the transition to most accurate perceptions of the system
 is solely a function of the measurement efficiency.. 
The figures show that a high knowledge of the state of the system is gained for $\eta \sim 0$ as $\eta$ increases.
This gain decreases for larger values of $\eta$. This observation is confirmed by explicit computation using the relative entropy, which satisfies $\frac{d }{d\eta} \left\langle  S\left( \rhog{t}  || \rhob{t}  \right) \right\rangle_\mathcal{B}  = \log \left( \frac{1-\sqrt{\eta}}{1+\sqrt{\eta}}\right)/(4\eta^{3/2})$. Thus, its rate of change and the information gain diverges for $\eta \rightarrow 0$ as a power law $\frac{d}{d\eta} \left\langle  S\left( \rhog{t}  || \rhob{t}  \right) \right\rangle_\mathcal{B} =-(1/6+1/2\eta)+\mathcal{O}(\eta^2)$, while it becomes essentially constant for intermediate values of $\eta$. In the transition to 
  most accurate perception 
the effective description  of the system changes from a mixed to a pure state, and the information gain becomes divergent as well as $\eta \rightarrow 1$.

\

\section{Discussion}

Different levels of information of a system amount to different effective descriptions. We studied these different descriptions for the case of a system being monitored by an observer, and compared this agent's description to that of other agents with a restricted access to the measurement outcomes. 
With continuous measurements as illustrative case study, we put bounds on the average trace distance between states that different agents assign to the system, and obtained exact results for the average quantum relative entropy.
 The expressions solely involve the state assigned by the less-knowledgeable agent, providing estimates for the distance to the exact state that can be calculated by the agent without knowledge of the latter.

The setting we presented here has a natural application to the case of a system interacting with an environment. 
For all practical purposes, one can view the effect of an environment as effectively monitoring the system with which it interacts~\cite{SchlosshauerRevModPhys2005,
zurekNatPhys2009}. Without access to the environmental degrees of freedom, the master equation that governs the state of the system takes a Lindblad form with Hermitian operators, as in Eq.~\eqref{eq:masterAlice}. However, access to the degrees of freedom of the environment can provide information of the state of the system, effectively leading to a dynamics governed by Eq.~\eqref{eq:master-filtered}. Access to a high fraction of the environment leads to a dynamics as in Eq.~\eqref{eq:master}, providing complete description of the state of the system by conditioning on the observed state of the environmental degrees of freedom. With this in mind, our results shed light on how much one can improve the description of a given system by incorporating information encoded in an environment~\cite{zurekNatPhys2009,ZwolakPRA2009,RiedelNJP2012,ZwolakSciRep2013,Brandao15,HorodeckiPRA2015,Olaya-CastroPRL2019}, as experimentally explored in \cite{PaternostroPRA2018,Jian-Wei-PanScienceBulletin2019}.
Note that since our bounds depend on the state assigned by the agent with less information, the above is independent of the unraveling chosen. It would also be interesting to extend our results and the connections to the dynamics of open systems to more general monitoring dynamics (e.g., non Hermitian operators or other noise models).

As brought up by an analysis  of a continuously-monitored harmonic oscillator,  
a large gain of information about the state of the system occurs when an agent has access to a small fraction of the measurement output,   
both when quantified by the trace distance and by the relative entropy. 
Our results thus complement the Quantum Darwinism program and related approaches~\cite{zurekNatPhys2009,ZwolakPRA2009,RiedelNJP2012,ZwolakSciRep2013,Brandao15,HorodeckiPRA2015,Olaya-CastroPRL2019}, where the authors  
compare the state of a system interacting with an environment and the state of fractions of such environment. 
While those works focused on the correlation buildup between the system and the environment, 
we instead address the subjective description that observers assign to the state of the system, conditioned on the information encoded in a given measurement record.

\
 
\noindent \emph{Acknowledgements.--- } This work was funded by the John Templeton Foundation, UMass Boston (project P20150000029279), and DOE grant DE-SC0019515.

\bibliography{referencesmultipleobservers}


\newpage 
\clearpage

\section*{Appendix}


\subsection{Derivation of bounds to average trace distance}

Using Eqs.~\eqref{eq:lindblad} and~\eqref{eq:masterAlice} in the main text and that $\rhog{0} = \rhoa{0}$, 
%
we find 
\begin{align}
\left\langle 1 - \tr{\rhog{T} \rhoa{T}} \right\rangle&=  - \left\langle \int_{\tr{\rhog{0} \rhoa{0}}}^{\tr{\rhog{T} \rhoa{T}}} d \tr{\rhog{t} \rhoa{t}} \right\rangle  = -\int_{F_0}^{F_T}    d \tr{\rhoa{t} \rhoa{t}}   = - 2 \int_{0}^{T}  \tr{ \rhoa{t} \deph\left[ \rhoa{t} \right]} dt \nonumber \\
&\!\!\!\!\!\!\!\!\!\!\!\!\!\!\!\!= +2\sum_\alpha \frac{1}{8 \tau_m^\alpha} \int_{0}^{T}  \tr{ \left[A_\alpha, \left[A_\alpha,\rhoa{t} \right] \right] \rhoa{t}} dt \nonumber \\
&\!\!\!\!\!\!\!\!\!\!\!\!\!\!\!\!=\sum_\alpha \frac{1}{4 \tau_m^\alpha} \int_{0}^{T}  \tr{ \left[\rhoa{t},A_\alpha \right] \left[A_\alpha,\rhoa{t} \right] } dt. 
\end{align}
This identity can be conveniently expressed in terms of the 2-norm of the commutator $[\rhoa{t},A]$ as
\begin{align}
\left\langle 1 - \tr{\rhog{T} \rhoa{T}} \right\rangle&=\sum_\alpha \frac{1}{4 \tau_m^\alpha} \int_{0}^{T}  \left\| \left[\rhoa{t},A_\alpha \right] \right\|_2^2 dt  = \sum_\alpha \frac{T}{4\tau_m^\alpha } \overline{\left\| \left[\rhoa{t},A_\alpha \right]   \right\|_2^2},
\end{align}
where we denote the time-average of a function $f$ by $\overline{f} \equiv  \int_0^T f(t) dt/T$. Note that the 
expression
$\sum_\alpha \frac{1}{4\tau_m^\alpha } \overline{\left\| \left[\rhoa{t},A_\alpha \right]   \right\|_2^2}$ plays the role of a time-averaged decoherence time  \cite{Chenu17,Beau17}, generalizing Eq. (11) in the main text.

This sets alternative bounds on the average distance between the state $\rhoa{t}$ assigned by $\mathcal{A}$ and the actual state of the system $\rhog{t}$, in terms of the effect of the Lindblad dephasing term acting on the incomplete-knowledge state $\rhoa{t}$,
\begin{align}
& T \sum_\alpha \tfrac{1}{4\tau_m^\alpha } \overline{\left\| \left[\rhoa{t},A_\alpha \right]   \right\|_2^2}   \le \left\langle \dist \left(\rhog{T},\rhoa{T}\right) \right\rangle \le  \sqrt{ T \sum_\alpha \tfrac{1}{4\tau_m^\alpha } \overline{\left\| \left[\rhoa{t},A_\alpha \right]   \right\|_2^2} } \, .    \nonumber
\end{align}

A short time analysis provides a sense of the evolution of the upper and lower bounds on the trace distance and how they compare to its variance.  
To leading order in a Taylor series expansion, 
\begin{align}
\mathcal{P}\left( \rhoa{\tau} \right)  & \approx 1+ 2 \tr{\rhoa{0} \deph \left[ \rhoa{0} \right] } \tau  = 1 - \sum_\alpha \frac{1}{4 \tau_m^\alpha} \tr{ \left[\rhoa{0},A_\alpha \right] \left[A_\alpha,\rhoa{0} \right] }  \tau,
\end{align}
and one finds 
\begin{align}
\label{eq-app:bounds-shorttime}
&\tau \sum_\alpha \tfrac{1}{4\tau_m^\alpha }  \left\| \left[\rhoa{0},A_\alpha \right]   \right\|_2^2   \le \left\langle \dist \left(\rhog{\tau},\rhoa{\tau}\right) \right\rangle \le  \sqrt{ \tau \sum_\alpha \tfrac{1}{4\tau_m^\alpha }  \left\| \left[\rhoa{0},A_\alpha \right]   \right\|_2^2 }.
\end{align}
Note that the behaviour of the trace distance is determined by the timescale in which decoherence occurs.

Using Eq.~(9) in the main text and Jensen's inequality one gets
\begin{align}
\left\langle \dist^2\left(\rhog{T},\rhoa{T}\right) \right\rangle \le  1 -\purityA,
\end{align}
which implies that the variance $\Delta \dist_T^2 \equiv \left\langle \dist^2\left(\rhog{T},\rhoa{T}\right) \right\rangle - \left\langle \dist\left(\rhog{T},\rhoa{T}\right) \right\rangle^2 $ satisfies
\begin{align}
\label{eq-app:boundvariance}
\Delta \dist_T^2 \le \purityA -\purityA^2.
\end{align}
In the short time limit this becomes
\begin{align}
\label{eq-app:bound-variance-shorttime}
\Delta \dist_\tau^2 \le -2 \tr{\rhoa{0} \deph \left[ \rhoa{0} \right] } \tau.
\end{align}

\subsection{Derivation of the average and variance of the quantum relative entropy}

Using that $\rhog{t}$ is pure, and that the von Neumann entropy is given by $S\left( \rho \right) \equiv - \tr{\rho \log \rho }$, we obtain that the average over the results unknown to agent $\mathcal{A}$ satisfy
\begin{align}
\label{eq:bound-relentropy}
\left\langle  S\left( \rhog{t}  || \rhoa{t}  \right) \right\rangle &= \left\langle   \tr{\rhog{t} \log \rhog{t}} \right\rangle - \left\langle   \tr{\rhog{t} \log \rhoa{t}} \right\rangle  = 0 -  \tr{\rhoa{t} \log \rhoa{t}}  = S\left(\rhoa{t}\right).
\end{align}
This sets a direct connection between the average error induced by assigning state $\rhoa{t}$ instead of the exact state $\rhog{t}$, as quantified by the relative entropy, in terms of the von Neumann entropy of the state accessible to agent $\mathcal{A}$.

In turn, the variance of the relative entropy satisfies
\begin{align}
\label{eq-app:bound-Var-relentropy}
\Delta S^2\left( \rhog{t}  || \rhoa{t}  \right) &= 
\left\langle  S^2\left( \rhog{t}  || \rhoa{t}  \right) \right\rangle -\left\langle  S\left( \rhog{t}  || \rhoa{t}  \right) \right\rangle^2  = \left\langle   \tr{\rhog{t} \log \rhoa{t}}^2 \right\rangle - S^2\left(\rhoa{t} \right) \nonumber \\
&\leq \left\langle  \tr{\rhog{t}}  \tr{\rhog{t} \log^2 \rhoa{t}}  \right\rangle  - S^2\left(\rhoa{t}\right)  =  \tr{\rhoa{t} \log^2 \rhoa{t}}  - S^2\left(\rhoa{t}\right),
\end{align}
using the Cauchy-Schwarz inequality in the third line. 
Note that this expression is identical to the variance of the operator $\left(-\log{\rhoa{t}}\right)$, which can be thought of the quantum extension to the notion of the `information content' or `surprisal' $\left(-\log p\right)$ in classical information theory.

\subsection{Bounds to the difference between perceptions of multiple agents}

Consider two agents $\mathcal{A}$ and $\mathcal{B}$ who simultaneously monitor different observables on a system. Each one has access to the measurement outcomes of their devices, but not to the results obtained by the other agent. The states $\rhoa{T}$ and $\rhob{T}$ that $\mathcal{A}$ and $\mathcal{B}$ assign to the system differ from the actual pure state $\rhog{T}$ that corresponds to the complete description of the system. For simplicity let us consider that $\mathcal{A}$  monitors a single observable $A$ and $\mathcal{B}$ monitors a single observable $B$.
The complete-description state of the system assigned by all-knowing agent $\mathcal{O}$ evolves according to
\begin{align}
\label{eq:master-twoobs}
d \rhog{t} = L\left[ \rhog{t} \right]dt + I_\mathcal{A}\left[ \rhog{t}\right] dW_t^\mathcal{A} + I_\mathcal{B}\left[ \rhog{t}\right] dW_t^\mathcal{B},
\end{align}
with the Lindbladian $L\left[ \rhog{t} \right] \nobreak\equiv\nobreak  -i\left[ H,\rhog{t} \right]  \nobreak +\nobreak  \deph_\mathcal{A} \left[ \rhog{t} \right] +  \deph_\mathcal{B} \left[ \rhog{t} \right] $, with  
 corresponding dephasing terms on observables $A$ and $B$.
The innovation terms $I_A$ and $I_B$ are defined as in Eq.~\eqref{eq:innovation} in the main text, 
 and $dW_t^\mathcal{A}$ and $dW_t^\mathcal{B}$ are independent noise terms.

The states of both observers satisfy
\begin{align}
d \rhoa{t} &= L\left[ \rhoa{t} \right]dt + I_{A}\left[ \rhoa{t}\right] dV_t^\mathcal{A} \\
d \rhob{t} &= L\left[ \rhob{t} \right]dt + I_{B}\left[ \rhob{t}\right] dV_t^\mathcal{B}.
\end{align}
Consistency between observers implies that their noises are related to the ones appearing in Eq.~\eqref{eq:master-twoobs} by~\cite{JacobsIntro2006,Bookjacobs2014}
\begin{align}
dW_t^\mathcal{A} &= \left( \tr{\rhoa{t} A} - \tr{\rhog{t} A} \right) \frac{dt}{\tau_m} + dV_t^\mathcal{A} \nonumber \\
dW_t^\mathcal{B} &= \left( \tr{\rhob{t} B} - \tr{\rhog{t} B} \right) \frac{dt}{\tau_m} + dV_t^\mathcal{B}.
\end{align}

As the state of each observer satisfies Eq.~\eqref{eq:bound-alice}, the triangle inequality provides the upper bound
\begin{align}
\label{eq:app-purity_to_tracedistanceABupperbound}
\big\langle \dist \left(\rhoa{T},\rhob{T}\right) \big\rangle_{\mathcal{A}\mathcal{B}} 
\le \sqrt{1  - \tr{\rhoa{T}^2} } + \sqrt{1  - \tr{\rhob{T}^2} },
\end{align}
and the lower bound
\begin{align}
\label{eq:app-purity_to_tracedistanceABlowerbound}
\big\langle \dist \left(\rhoa{T},\rhob{T}\right) \big\rangle_{\mathcal{A}\mathcal{B}}
\ge\left| \tr{\rhoa{T}^2} - \tr{\rhob{T}^2} \right|.
\end{align}

\color{black}
\subsection{Illustration --- evolution of limits to perception}

\begin{figure}
  \centering
        \includegraphics[trim=00 00 00 00,width=0.32\textwidth]{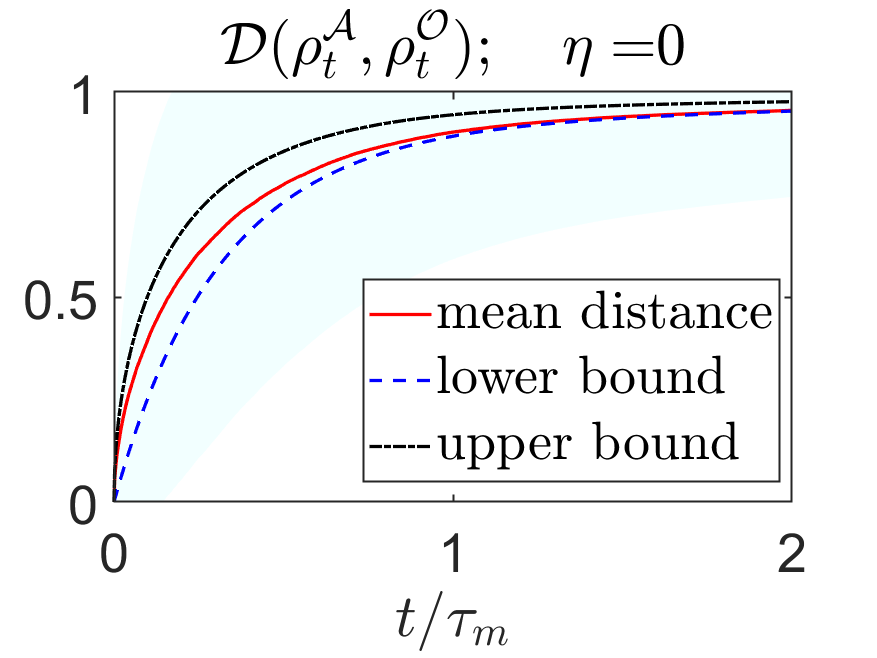}
         \includegraphics[trim=00 00 00 00,width=0.32\textwidth]{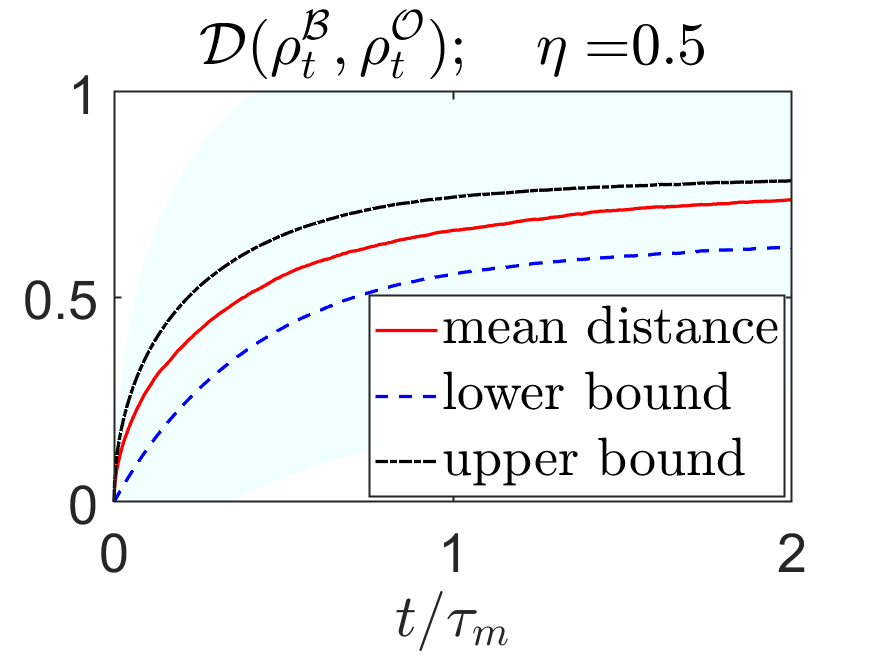}  
       \includegraphics[trim=00 00 00 00,width=0.32\textwidth]{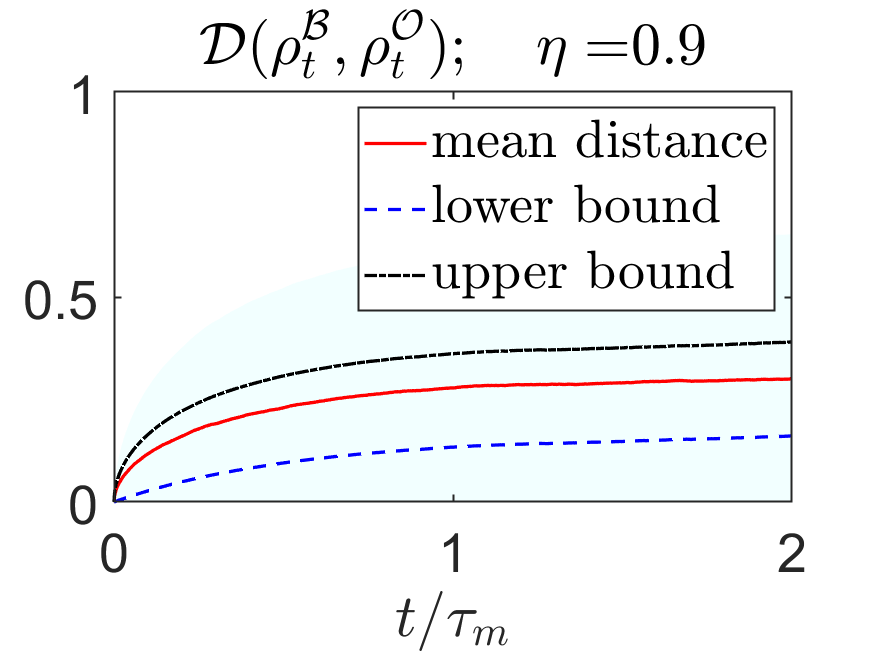}  
\caption{
\label{fig--app:evolutionBounds} 
{\bf 
Evolution of the average trace distance and its bounds.}
Simulated evolution of the average trace distance $\left\langle \dist\left( \rhog{T}, \rhob{T}  \right) \right\rangle$ between complete and incomplete descriptions 
for a spin chain initially in a paramagnetic state on which individual spin components $\sigma_j^z$ are monitored. 
The simulation corresponds to $N = 6$ spins, with couplings $J\tau_m = h\tau_m = 1/2$.
 The upper and lower bounds~\eqref{eq:bound-bob} on the average trace distance is depicted by dashed lines, while the shaded area represents the (one standard deviation) confidence region obtained from the upper bound~\eqref{eq:boundvariance}
 on the standard deviation in the main text, calculated with respect to the mean distance.
For $\eta = 0 $ (left), agent $\mathcal{A}$, without any access to the measurement outcomes, has the most incomplete description of the system.
After gaining access to partial measurement results, with $\eta = 0.5$ (center) $\mathcal{B}$ gets closer to the complete description of the state of the system.
Finally, when $\eta = 0.9$ (right),
access to enough information provides $\mathcal{B}$ with an almost complete description of the state. Importantly, in all cases the agent can bound how far the description possessed is from the complete one solely in terms solely of the purity $\purityB$.
}
\end{figure}

\begin{figure}
  \centering
        \includegraphics[trim=00 00 00 00,width=0.32\textwidth]{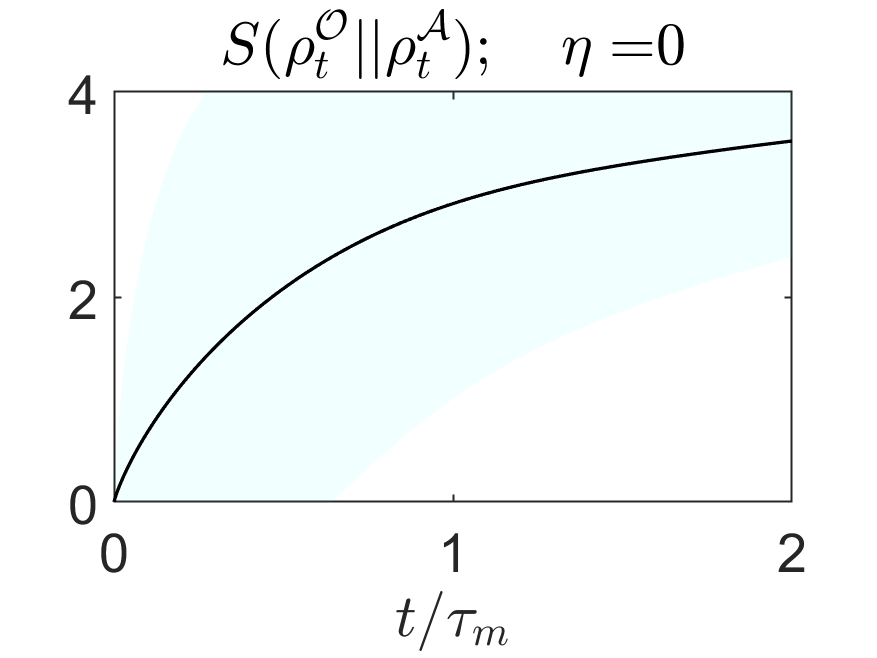}
         \includegraphics[trim=00 00 00 00,width=0.32\textwidth]{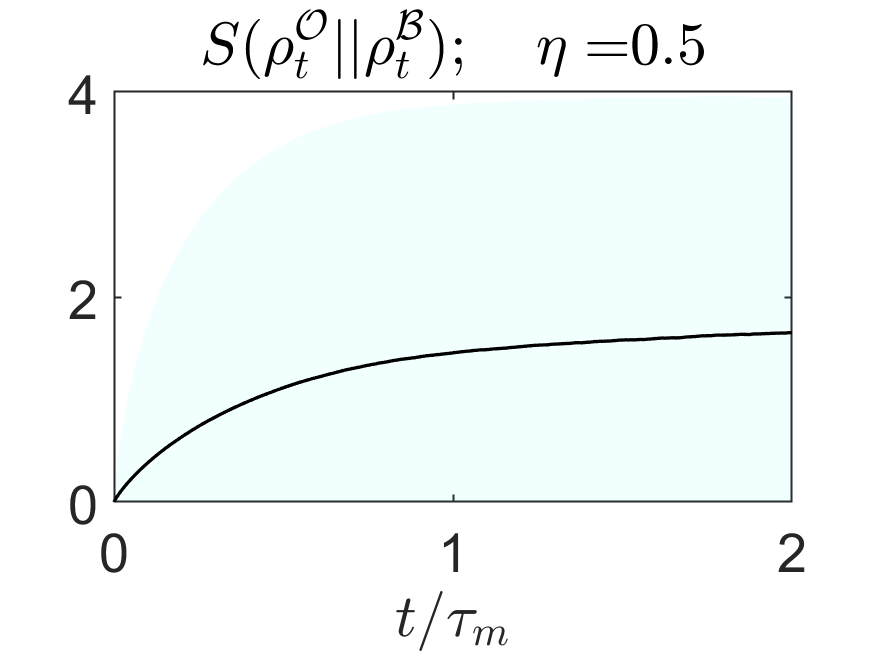}  
       \includegraphics[trim=00 00 00 00,width=0.32\textwidth]{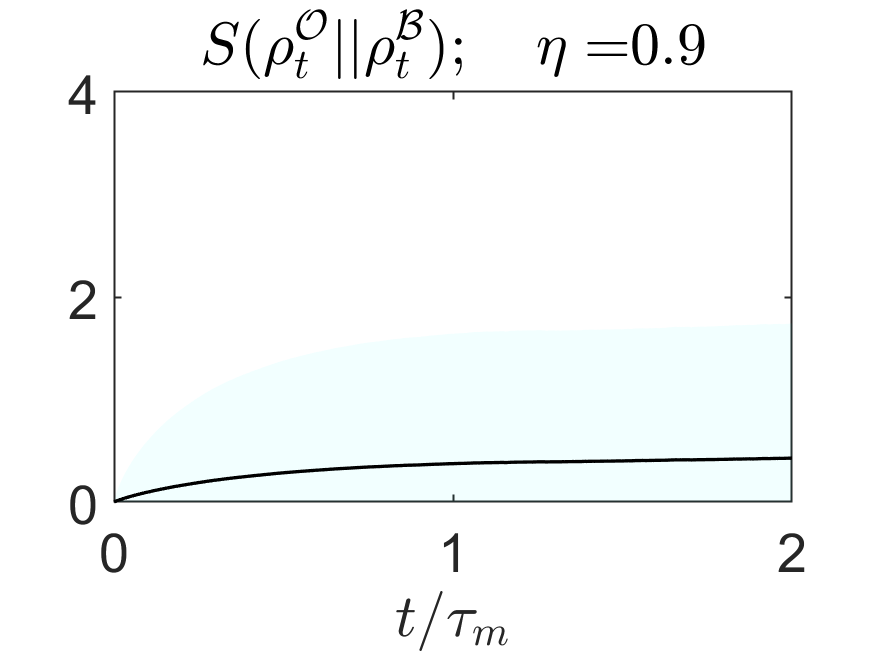}  
\caption{
\label{fig--app:evolutionEntropyBounds} 
{\bf 
Evolution of the average relative entropy and its bounds.}
Simulated evolution of the average relative entropy $\left\langle S\left( \rhog{T} || \rhob{T}  \right) \right\rangle$ between complete and incomplete descriptions for a spin chain on which the $z$ components of individual spins are monitored.
The shaded area represents the (one standard deviation) confidence region obtained from the upper bound 
 on the standard deviation of the relative entropy, Eq.~\eqref{eq:bound-Var-relentropy} in the main text.
As in the case of the trace distance, access to more information leads to a more accurate state assigned by the agent.
}
\end{figure}

We consider the case of observer $\mathcal{O}$ monitoring the spin components $\sigma_j^z$ on a 1D transverse field Ising model, with the Hamiltonian defined in Eq.~\eqref{eq:spinchainH} of the main text. 
Figure~\ref{fig--app:evolutionBounds} shows the evolution of the average trace distance $\left\langle \dist\left( \rhog{T}, \rhob{T}  \right) \right\rangle$ between the complete description and $\mathcal{B}$'s partial one, along with the bounds~\eqref{eq:bound-bob}, for different values of the monitoring efficiency $\eta$.
Figure~\ref{fig--app:evolutionEntropyBounds} shows the evolution of the average relative entropy $\left\langle S\left( \rhog{T} || \rhob{T}  \right) \right\rangle$.
The dynamics are simulated by implementation of the monitoring process as a sequence of weak measurements modeled by Kraus operators acting on the state of the system. Specifically, the evolution of $\rhog{t}$ and corresponding state $\rhob{t}$ with partial measurements is numerically obtained from assuming two independent measurement processes, as in~\cite{JacobsIntro2006}.


\subsection{Illustration --- transition to complete descriptions}

Consider the case of a one dimensional harmonic oscillator  with position and momentum operators $X$ and $P$. 
We assume agent $\mathcal{B}$ is monitoring the position of the harmonic oscillator, with an efficiency $\eta$. 
The dynamics of state $\rhob{t}$ is dictated by Eq.~\eqref{eq:master-filtered} in the main text  
for the case of a single monitored observable,
with 
\begin{align}
  \deph \left[ \rhob{t} \right] &=  \frac{1}{8 \tau_m } \left[X, \left[X,\rhob{t} \right] \right];  \qquad I_{X} \left[ \rhob{t}\right] =     \frac{1}{\sqrt{4\tau_m }} \left( \{ X, \rhob{t} \} - 2\tr{X \rhob{t}} \rhob{t} \right) .
\end{align}  
Such dynamics preserves the Gaussian property of states. For these, the variances 
\begin{align}
v_x &\equiv \tr{\rhob{t} X^2} - \tr{\rhob{t} X}^2, \\
v_p &\equiv \tr{\rhob{t} P^2} - \tr{\rhob{t} P}^2,
\end{align}
 and covariance 
 \begin{align}
 c_{xp} \equiv \tr{\rhob{t} \frac{\{X,P\}}{2}} -  \tr{\rhob{t} X} \tr{\rhob{t} P},
 \end{align}
 satisfy the following set of differential equations (in natural units)~\cite{DohertyPRA99,JacobsIntro2006}
\begin{subequations}
\label{eq:app-diffeqsGaussian}
\begin{align}
\frac{d}{dt}v_x &= 2 \omega c_{xp} - \frac{\eta}{\tau_m} v_x^2,      \\
\frac{d}{dt}v_p &= -2 \omega  c_{xp} + \frac{1}{4\tau_m} -\frac{\eta}{\tau_m} c_{xp}^2,      \\
\frac{d}{dt}c_{xp} &= \omega v_p -  \omega  v_x  -\frac{\eta}{\tau_m} v_x c_{xp}.
\end{align}
\end{subequations}
While first moments do evolve stochastically, the second moments above satisfy a set of deterministic coupled differential equations. This in turn implies that the purity of the state, which can be obtained from the covariance matrix~\cite{PhysRevA.68.012314,ParisGaussian2005,
HayashiGaussian2007,PirandolaRevModPhys2012,adesso2014continuous}
\begin{align} 
\sigma(t) \equiv 
\begin{bmatrix} 
v_x & c_{xp} \\
c_{xp} & v_p
\end{bmatrix}
\end{align}
as 
\begin{align}
\purityB = \frac{1}{2 \sqrt{ \det{ [\sigma(t) ]}}},
\end{align}
evolves deterministically as well.

The solution for long times can be derived from Eqs.~\eqref{eq:app-diffeqsGaussian}, giving
\begin{subequations}
\label{eq:app-solGaussian}
\begin{align}
c_{xp}^{ss} &=  - \frac{ \omega \tau_m \pm \sqrt{\omega^2 \tau_m^2 + \eta/4 } }{\eta},
 \\ v_x^{ss} &=  \sqrt{   \frac{2 \omega \tau_m}{ \eta }   c_{xp}^{ss}}\, ,  \\
v_p^{ss} &=   v_x^{ss}\left( 1+ \frac{  \eta  }{\omega \tau_m}  c_{xp}^{ss} \right) , 
\end{align}
\end{subequations}
which provides the long-time asymptotic  value of the purity as a function of the measurement efficiency.
The latter turns out to have the following simple expression
\begin{align}
\purityB &= \frac{1}{2\sqrt{v_x^{ss}v_p^{ss} - (c_{xp}^{ss})^2}}  = \frac{1}{2 \sqrt{  \frac{2 \omega \tau_m}{ \eta }   c_{xp}^{ss} \left( 1 +  \frac{  \eta  }{\omega \tau_m}  c_{xp}^{ss}   \right)  - (c_{xp}^{ss})^2  }}  = \frac{1}{2 \sqrt{  \frac{2 \omega \tau_m}{ \eta }   c_{xp}^{ss}   + (c_{xp}^{ss})^2  }} \nonumber \\
& = \frac{1}{2 \sqrt{  \frac{ \tau_m}{ \eta }   \left( \frac{1}{4\tau_m} -\frac{\eta}{\tau_m} (c_{xp}^{ss})^2  \right)  + (c_{xp}^{ss})^2  }}  = \frac{1}{2 \sqrt{  \frac{1}{4\eta} } }  = \sqrt{\eta}.
\label{PurEff}
\end{align}
Using that 
\begin{align}
\label{eq:boundDist-bob}
1  -  \purityB   \le &\left\langle \dist \left(\rhog{T},\rhob{T}\right) \right\rangle_\mathcal{B} \le \sqrt{1  - \purityB},
\end{align}
then implies
\begin{align}
1  -  \sqrt{\eta}   \le \left\langle \dist \left(\rhog{T},\rhob{T}\right) \right\rangle_\mathcal{B} \le \sqrt{1  - \sqrt{\eta} }.
\end{align}

The entropy of a 1-mode Gaussian state can be expressed in terms of the purity of the state as 
\begin{align}
S\left(\rhob{T} \right) &= \left( \frac{1}{2\purityB } + 1/2 \right) \log \left( \frac{1}{2\purityB } + 1/2 \right)  -    \left( \frac{1}{2\purityB } - 1/2 \right) \log \left( \frac{1}{2\purityB } - 1/2 \right) .
\end{align}
Then, using that $
\left\langle  S\left( \rhog{t}  || \rhob{t}  \right) \right\rangle_\mathcal{B} = S\left(\rhob{t}\right).
$ and Eq.~\eqref{PurEff}, we obtain that for long times,
\begin{align}
\left\langle  S\left( \rhog{t}  || \rhob{t}  \right) \right\rangle_\mathcal{B} &= S\left( \rhob{T}\right)  =  \left( \frac{1}{ 2\sqrt{\eta} } + \frac{1}{2} \right) \log \left( \frac{1}{2\sqrt{\eta} } + \frac{1}{2} \right)  -  \left( \frac{1}{ 2\sqrt{\eta} } - \frac{1}{2} \right) \log \left( \frac{1}{2\sqrt{\eta} } - \frac{1}{2} \right) .
\end{align}

\end{document}